# Hole Concentration Effect on the Microwave Nonlinearity of $Tl_2Ba_2CaCu_2O_{8\pm\delta}$ Superconducting Thin Films

**M.M. Bischak**[1], **J. Thomas**[1], **R.R. Philip**[2], **and S.K. Remillard**[1,3]
[1] Hope College, 27 Graves Place, Holland, MI 49423, USA.
[2] Thin Film Research Lab, UC College, Aluva, Cochin, Kerala, India.

E-mail: remillard@hope.edu

**Abstract.** The carrier concentration of $Tl_2Ba_2CaCu_2O_8$ films was modified by annealing in $N_2$ gas. X-ray analysis of the structure and the oxygen content revealed a correspondence between carrier concentration and oxygen depletion. The $T_C$ and nonlinear surface impedance was measured using a dielectric resonator and the nonlinearity slope parameter $r=\Delta X_S/\Delta R_S$ was found to converge to unity at the critical temperature, indicating a dominance of Josephson fluxon hysteresis on the nonlinearity. Highly inductive nonlinearity was observed in a small range of doping levels between $0.180<p<0.195$ holes/Cu, which does not include the optimal doping level of 0.16 holes/Cu.

## 1. Introduction
High temperature superconductor (HTS) thin films are used in passive microwave devices such as RF filters[1] and MRI pick-up coils[2]. These applications are best served by materials with little nonlinearity, which is manifest as a dependence of the surface impedance, $Z_S$, on the microwave magnetic field, $H_{RF}$, as well as with signal distortion[3]. In this experiment, modification of the hole doping moves the nonlinear response of the thin film to a new quiescent operating point, and permits two key observational opportunities. First is the opportunity to examine nonlinearity at lower power with suppressed carrier concentration as if it were being operated at higher power. Second is the opportunity to directly examine the effect of carrier concentration on the nonlinearity of $Z_s$.

The starting material for these experiments comes from a large supply of 50 mm diameter $Tl_2Ba_2CaCu_2O_8$ thin films on $LaAlO_3$ substrates, which presents a unique opportunity to conduct these experiments that may otherwise be prohibited by material supply limitations.

## 2. Microwave Nonlinearity in $Tl_2Ba_2CaCu_2O_8$ Films
A temperature and $H_{RF}$ dependent fit function for $Z_S$ of $Tl_2Ba_2CaCu_2O_8$ was used by Gaganidze, *et al.*[4], with various terms corresponding to the mechanisms behind the loss. At very low $H_{RF}$, weak links formed by twin and grain boundaries are in the Meissner state and $Z_S$ is dominated by leakage through the weak links. $R_S$ then primarily depends on $H_{RF}^2$ [5], although at higher $H_{RF}$ the grain boundaries decouple and the dependence becomes weaker. A low field saturating flux-flow contribution varies as $(1+H_0/H_{RF})^{-1/2}$, where the characteristic scaling field $H_0$ can be especially pronounced in $Tl_2Ba_2CaCu_2O_8$ which has very weak links[6]. Above $J_{Cj}$ Josephson fluxons nucleate and their dissipation is hysteretic, mostly due to pinning, as characterized by $R_S \propto H_{RF}$ and $X_S \propto H_{RF}$[7]. Although stronger $H_{RF}$ dependence is seen if the fields are enhanced at film edges [7] or at pores and exposed cleavage plane edges, $H_{RF}^2$ dependence can be attributed to leakage current through weak links[8]. Oxygen deficiency causes enhanced kinetic inductance in weak links in YBCO[9], and other experiments have also found grain boundary transport[10] and pinning[11] to be influenced by doping.

---
[3] To whom any correspondence should be addressed.

The dimensionless slope parameter ($r=\Delta X_S/\Delta R_S$) of $r \approx 1$ along with $R_S \propto H_{rf}$, or $H_{rf}^2$ in the case of edge effects, indicates hysteresis loss due to the nucleation, annihilation and pinning of fluxons[4]. However, $r >> 1$ corresponds to intrinsic nonlinearity[8], which scales as the $H_{C1}$ of the bulk material. In the case of unaligned small grains, the weakly coupled grain model has been shown to be consistent with an observed negative value of $r$ at high reduced temperature[12]. In these measurements $H_{RF}$ remains under about 1,500 A/m and the reduced temperature remains under $t=0.95$, meaning that $H_{C1}$ is not approached. Thus the nonlinearities in these measurements do not involve the intrinsic effects of Abrikosov fluxons.

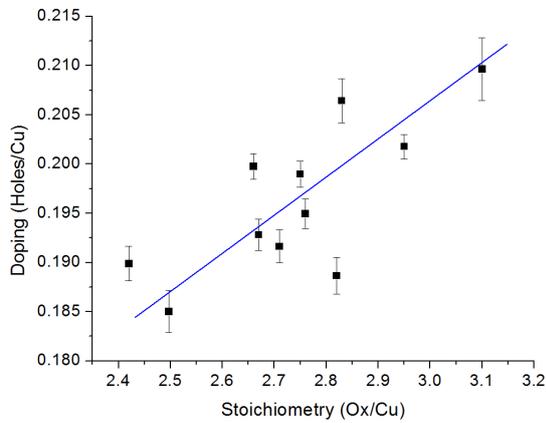
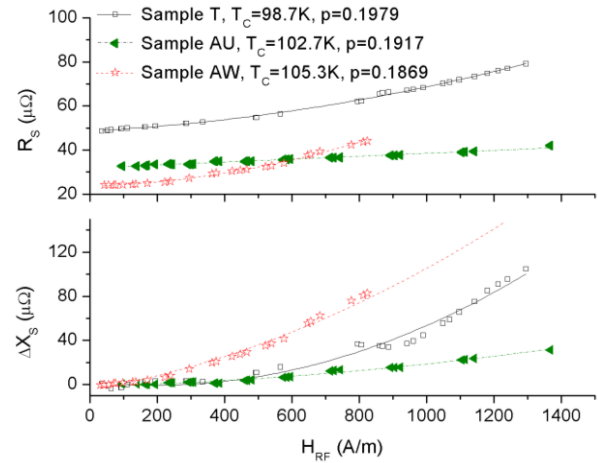

**Figure 1.** Doping from $T_C$ as a function of Oxygen/Copper ratio determined from EDS.

**Figure 2.** $R_S$ and $\Delta X_S$ for films of three doping levels at 5.5 GHz and reduced temperature of 0.79 with 2$^{nd}$ order polynomial fits.

### 3. Experiment
3.1. The Films and Their Hole Doping
The films were made by DuPont Superconductivity by annealing off-axis magnetron sputtered amorphous BaCaCuO films in a $Tl_2O_3$ vapor at 1 atm[13]. We confirmed the manufacturer's report that $R_S$ at 8.1 GHz and 77K was less than 200μΩ and that the thicknesses were about $d=400$ nm.

Carrier density, $p$, was depleted by annealing for 7 hours in $N_2$ flowing at 20 cm$^3$/min between 250°C and 360°C. $p$ is related to $t=T_C/T_{C,max}$ through the phase boundary parabola[14] $p=0.16 \pm \sqrt{(1-t)/82.6}$ where $T_{C,max}$ is about 112 K for $Tl_2Ba_2CaCu_2O_8$. The positive sign corresponds to overdoping, and the 2212 phase is over-doped. So $N_2$ annealing can modify the $T_C$ [15]. Tl becomes mobile only above our anneal temperatures. [16]. Thus, changes in $T_C$ correspond to changes in oxygen. Figure 1 shows holes per Cu, from $T_C$, plotted against oxygen per Cu, from energy dispersive X-ray spectroscopy. A variation in oxygen results in a scattered variation in doping illustrating the connection between oxygen depletion and carrier concentration.

3.2. Surface Impedance Measurement
The surface impedance in Figure 2 is measured using a polished sapphire/HTS resonator following a well-established 3-sample round-robin [17,18]. $X_S(H_{RF})$ is also measured in the round robin, and corresponds to a variation in effective penetration depth, $\lambda$, as $\Delta X_S=(\omega\mu_o)\Delta\lambda$, which comes directly from the resonant frequency, $f_o$, by $\Delta X_S = 2G \cdot (\Delta f / f_o)$. $G$ is the sample geometry factor, in our case 203 $\Omega$, and $\Delta f$ is the resonant frequency shift.

Thin films require a correction to $R_S$ for the substrate [19]. Surface reactance, also corrected for the film thickness, is $\Delta X_S = \Delta X_{S,eff} \tanh(d/\lambda)$ where $\Delta X_{S,eff}$ is computed from the frequency shift. The effective penetration depth, $\lambda$, is a composite of the bulk London depth and field penetration at weak links. Iteration is used to find $\lambda$ using $\lambda_{ab}(0) = 200$ nm [20] applied to the d-wave dirty limit $\lambda(T) = \lambda(0)/\sqrt{1-(T/T_C)^2}$ which is applicable to these DuPont $Tl_2Ba_2CaCu_2O_8$ thin films [21].

## 4. Results

$R_S$ and $\Delta X_S$ of three representative films are shown at t=0.79 in Figure 2. All films exhibit both linear and quadratic terms in $R_S(H_{RF})$, although all films with $T_C$ around 102 K have a much weaker function of $Z_S(H_{RF})$. In all cases, the $X_S$ has both linear and quadratic field dependence. For these films, all examined at 77K and above, including the flux flow term proportional to $(1 + H_0/H_{RF})^{-1/2}$ did not improve the fit. Using a wider temperature range, Hein *et al.*[22] scaled the currents at different temperatures and found the $H_{RF}$ dependence of $R_S$ to be independent of Oxygen doping in YBCO.

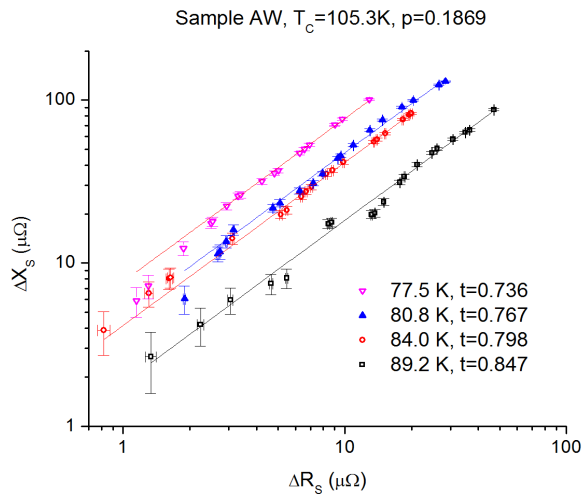
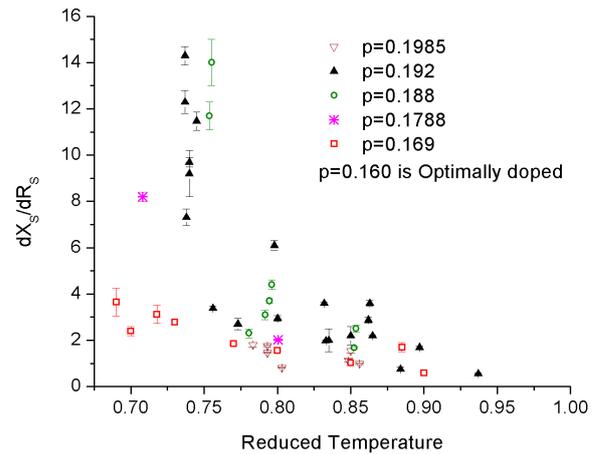

**Figure 3.** $X_S$ varies nearly linearly with $R_S$, with $H_{RF}$ as the implicit variable. The slope (corresponding to height on the log graph) decreases with increasing T from 77.5K.

**Figure 4.** All slopes of 20 samples fall in a band of values that converges to unity at $T_C$. Binning according to doping level reveals the trends.

$X_S$ and $R_S$ usually have similar dependence on $H_{RF}$ [23], thus producing a linear variation of $\Delta X_S$ with $\Delta R_S$, with $H_{RF}$ as the implicit parameter, shown in Figure 3. The slope, $r = \Delta X_S/\Delta R_S$, varies with doping, $p$, and with reduced temperature, $t$. Forty five TBCCO/LAO wafers were used in some way throughout this study and the temperature dependence of the slopes fall within a region that converges at unity at $T_C$ and ranges from $r=2$ to $r=15$ at $t=0.75$.

The dependence of r on $p$ at constant $t$ appears to be quite scattered. But by viewing the slope results as in Figure 4, with the slopes binned according to $p$, the trend stands out. For doping in the range $0.180 < p < 0.195$ the slope becomes large ($r > 10$) at about $t = 0.75$. For films in this range of carrier concentration, the low temperature nonlinearity is dominated by the kinetic inductance. Thus the low temperature inductive nonlinearity survives to the highest temperature for this doping region. The dramatic increase in r does not occur above and below this doping range. For p<0.180 and p>0.195, the slope for $0.75<t<1$ remains in the range of only $1<r<4$ and the surface resistance is linear in $H_{RF}$, indicating impedance nonlinearity dominated by Josephson fluxon hysteresis.

## 5. Conclusion

The doping of 400 nm thick $Tl_2Ba_2CaCu_2O_8$ films on $LaAlO_3$ substrates was modified by annealing the films in $N_2$ gas below 400°C. The nonlinear slope parameter was found to peak for carrier concentrations in the range of $0.180<p<0.195$ holes/Cu indicating the dominance of significant inductive loss in weak links to higher temperatures. Outside this doping range, hysteretic loss from nucleation, annihilation and pinning of Josephson fluxons dominates the nonlinear surface impedance. In continuing work, these results are being applied to the preparation of patterned single resonators in order to investigate the influence of carrier concentration on intermodulation distortion.

## 6. Acknowledgement

This work was supported by grants DMR/1206149 and PHY-DMR/1004811 from the U.S. National Science Foundation, and by a grant from the Research Corporation for Science Advancement. RRP acknowledges UBCHEA, NY for their support through the United Board Fellows program.